\documentclass[preprint,12pt]{aastex}
\usepackage{epsfig, amsmath, amsfonts}

\newcommand\Pm{ {\rm Pm}}

\def\beq{ \begin{equation} }
\def\eeq{ \end{equation} }
\def\spose#1{\hbox to 0pt{#1\hss}}  %from Scott Tremaine
\def\ltsim{\mathrel{\spose{\lower.5ex\hbox{$\mathchar"218$}}
\raise.4ex\hbox{$\mathchar"13C$}}}
\def\gtsim{\mathrel{\spose{\lower.5ex\hbox{$\mathchar"218$}}
\raise.4ex\hbox{$>$}}}

\slugcomment{Version: \today}

\begin{document}

\title{\bf\LARGE On the Magnetic Prandtl Number Behavior\\
of Accretion  Disks}
\author{ Steven A. Balbus\altaffilmark{1,2}}
\author{ Pierre Henri  \altaffilmark{1}}
%\author{ Christopher Reynolds  \altaffilmark{2}}

\altaffiltext{1}{Laboratoire de Radioastronomie, \'Ecole Normale
Sup\'erieure, 24 rue Lhomond, 75231 Paris CEDEX 05, France
  \texttt{steven.balbus@lra.ens.fr}}

  \altaffiltext{2}{Adjunct Professor, Department of Astronomy, University
  of Virginia, Charlottesville, VA 22903}

\begin{abstract}
We investigate the behavior of the magnetic Prandtl number (ratio
of microscopic viscosity to resistivity) for accretion sources.
Generally this number is very small in standard accretion disk models,
but can become larger than unity within $\sim 50$ Schwarzschild radii
of the central mass.  Recent numerical investigations suggest a marked
dependence of the level of MHD turbulence on the value of the Prandtl
number.  Hence, black hole and neutron star accretors, i.e. compact X-ray
sources, are affected.  The astrophysical consequences of this could be
significant, including a possible route to understanding the mysterious
state changes that have long characterized these sources.

\end{abstract}


\maketitle

\section{Introduction}

Magnetohydrodynamic (MHD) turbulence differs from ordinary hydrodynamic
turbulence in at least one very important respect: whereas the latter
generally has only one dissipative scale (viscous), MHD turbulence has
two (viscous and resistive).  This raises the question of whether the
classical Kolmogorov picture, in which the large scale energetics of the
turbulenct cascade is essentially independent of small scale dissipation
physics, remains valid in MHD turbulence.  And if it is not valid,
what are some possible astrophysical consequences?

The dimensionless ratio of the kinematic viscosity $\nu$ to the electrical
resistivity $\eta$ is known as the magnetic Prandtl number, $\Pm$.
Balbus \& Hawley (1998) suggested that even if both the viscous and
resistive dissipation scales are very small, the saturation level of the
MHD turbulence produced by the magnetorotational instability (MRI) should
be sensitive to $\Pm$, at least in the neighborhood of $\Pm\simeq 1$.
Their argument was as follows.  If $\Pm \ll 1$, the resistive scale
is much larger than the viscous scale.  Assuming that the velocity
fluctuations do not greatly exceed the Alfv\'enic fluctuations at the
resistive scale, viscous stresses on the resistive scale would be small.
This would mean that even if relatively large velocity gradients
accompanied magnetic dissipation, these gradients would not produce
stresses that would interfere with the dynamics of the field reconnection.
Since Lorentz forces drive the MRI, the dissipation of the magnetic
field is an important regulatory mechanism for the saturation level
of the turbulence.  On the other hand, if $\Pm \gg 1$, and the viscous
scale were significantly larger than the resistive, then the resulting
dynamical stresses would likely be relatively large when the magnetic
field is dissipated at the small resistive scale.  (This assumes that
significant velocity fluctuations accompany small scale reconnection.
Such fluctuations would be heavily damped at the resistive scale,
and any reconnection would have to be very slow.)  These stresses would
then interfere with the field reconnection and dissipation, leading to a
build-up of magnetic energy that cascades upwards, back to larger scales
(Brandenburg 2001).

Balbus \& Hawley (1998) were motivated by the possibility that the
turbulent properties of accretion disks might be different in the regimes
$\Pm \gg 1$ and $\Pm \ll 1$.  At the time, direct numerical simulation of
flows with different Prandtl numbers was very difficult, and these authors
attempted only the crudest of tests by varying the level of artificial
viscosity in the ZEUS MHD code at fixed resolution.  These preliminary
experiments did, however, show a higher level of saturation for a larger
viscosity.  Since this is an example in which increasing a dissipation
coefficient actually raised the level of turbulent activity, it was a
noteworthy result.

A decade on, it is possible to do much better.  There is a definite
sensitivity to $\Pm$ in numerical simulations of MHD turbulence.  A $\Pm$
dependence has been observed for a number of years now in stirred magnetic
turbulence (Schekochihin et al. 2004, 2005).  For example, fluctuation
dynamos at large Pm were found numerically by Schekochihin et al. (2004),
but until recently there was some question as to whether a low $\Pm$
fluctuation dynamo even existed; this has now been answered affirmatively
(Iskakov et al. 2007; Schekochihin et al. 2007).  For astrophysical
accretion flows, MRI calculations are of direct interest, and the last
year has seen the first $\Pm$ studies in ``shearing box'' simulations.
Zero mean field calculations have been carried out by Fromang et
al. (2007), while Lesur \& Longaretti (2007) studied a mean vertical
field.  In the latter investigation, the radial-azimuthal component
of the stress tensor behaved linearly over the range $0.12 < \Pm < 8$
with no apparent sign of approaching an asymptote (Iskakov et al. [2007]
did, however, appear to be reaching saturation levels in some of their
driven turbulence runs).  With extensive numerical evidence of a $\Pm$
dependence in MHD turbulence, a natural question to raise is what is the
behavior of $\Pm$ in classical accretion disk models?  In particular,
is there a transition from $\Pm \ll 1$ to $\Pm \gg 1$ in phenomenological
models that have been used to model AGN and compact X-ray sources?

In this paper, we examine the magnetic Prandtl number behavior of
classical $\alpha$ models.  In fact, the only feature of these models that
is important for our purposes is that the free energy of differential
rotation be locally dissipated---a variable $\alpha$ parameter, for
example, would hardly change our conclusions at all.  Throughout the
regime of interest, the disk is fully ionized and collision dominated (see
\S 2 below), so that the Spitzer (1962) values for the resistivity and
viscosity are appropriate.  Our principal finding is that generally $\Pm
\ll 1$ nearly everywhere in classical $\alpha$ models, with one robust
and important exception: on scales less than $\sim 100$ Schwarzschild
radii in black hole and neutron star disks.  It is extremely tempting
to associate this Prandtl number transition with a physical transition
in the properties of the accretion flow, here motivated by ``first
principle'' physics.  Further discussion of this point is presented below.

An outline of the paper is as follows.  \S 2 presents preliminary
estimates of important parameter regimes.  \S 3 is the heart of the paper, in
which we calculate the behavior of $\Pm$ in classical $\alpha$ disk models.
Transitions from low $\Pm$ to high $Pm$ regions occur only in disks
around black holes and neutron stars.  Finally, \S 4 is a discussion
of the possible astrophysical consequences of having both high $\Pm$
and low $\Pm$ regions in the same disk.  It is argued that high and
low X-ray states (e.g. McClintock \& Remillard 2006)
may be related to an unstable interface between $\Pm<1$
and $\Pm>1$ regions of the disk.

\section {Preliminaries}

The magnetic Prandtl number is not a standard parameter of accretion
theory, so let us begin with a brief orientation in the temperature-density
parameter space.  Throughout this work, the fiducial disk plasma is taken
to be a mixture
of 90\% hydrogen and 10\% helium (by number).   Following the discussion
in Spitzer (pp. 138-9), we estimate an averaged resistivity of such a fully ionized
gas as 
\beq
\eta = {5.55\times 10^{11}\ln\Lambda_{\rm e\,H}\over  T^{3/2}}\ {\rm cm}^2\ {\rm s}^{-1},
\eeq
where $T$ is the temperature in Kelvins, and $\Lambda_{\rm e\,H}$ is the Coulomb
logarithm for electron-proton scattering.  (Modifications in the
logarithm due to electron-helium
scattering, here a minor effect, are ignored.)  

The kinematic viscosity of the same gas is estimated to be
\beq
\nu = {1.6 \times 10^{-15} T^{5/2} \over \rho\ln\Lambda_{\rm H\,H}}\ {\rm cm}^2\ {\rm s}^{-1},
\eeq
where $\rho$ is the mass density and $\ln\Lambda_{\rm H\,H}$ is the 
Coulomb logarithm for scattering of protons by protons.  (See Appendix for a derivation of
these results and a discussion of the Coulomb logarithms.)
This gives a magnetic Prandtl number of
\beq\label{PPm}
\Pm=2.9\times  10^{-27} {T^4\over\rho\ln\Lambda_{\rm e\,H}\ln
\Lambda_{\rm H\,H}}.
\eeq
The two logarithms differ from one another for temperatures in 
excess of $4.2\times10^5$ K (see Appendix).  If $l$ is the product of
the two Coulomb logarithms normalized to a nominal value of 40,
\beq
\Pm=\left(T\over 4.2\times 10^6\, {\rm
K} \right)^4 \left(10^{22}\, {\rm cm}^{-3}\over l\  n_H\right)
=
\left(T\over 4.2\times 10^4\, {\rm
K} \right)^4 \left(10^{14}\, {\rm cm}^{-3}\over  l\   n_H\right),
\eeq
where $n_H$ is the number density of hydrogen atoms.  The two last
forms that are given for $\Pm$ are convenient for applications to a $10\,
M_\odot$ (binary) and $10^8\, M_\odot$ (AGN) black hole, respectively.

Finally, it is required to justify quantitatively
the statement in the Introduction
that the disk plasma is collisional near the transtion point $\Pm=1$.
We shall refer to a plasma as ``dilute'' (as opposed to collisional)
if the product of the ion cyclotron frequency $\omega_{ci}$ and the
ion-ion collision time $t_{ci}$ is greater than one.  The proton cyclotron
frequency may be written
\beq
\omega_{ci}= 8.6\times 10^{-4}\left( n_H T\over\beta\right)^{1/2},
\eeq
where we have introduced the plasma $\beta$ parameter, the ratio of
the gas to magnetic pressure.  For a gas of cosmic abundances, 
\beq
\beta = 2.3 \left(8\pi n_HkT\over B^2\right).
\eeq
In a dilute plasma, it is not appropriate to use the Spitzer (1962)
form of the viscosity, as we have done above, hence we need a numerical
estimate of $\omega_{ci}t_{ci}$.  (It is also not strictly correct to use
the Spitzer resistivity, but the correction here is relatively minor.)
Following the prescription set forth in the Appendix (divide the nominal
proton-proton collision time by a factor of 1.5 to include the effects
of proton-helium collisons), we obtain
\beq
\left(\omega_{ci} t_{ci}\right)^2 \simeq 9\times 10^{-7} {T^4\over n_H\beta}.
\eeq
where we have taken the relevant Coulomb logarithm to be 7.  This should
be compared directly to $\Pm \simeq 3\times10^{-5}T^4/n_H$ from equation
(\ref{PPm}).  For a given value of $\beta$, the temperature and density
dependence of $\omega_{ci}^2t_{ci}^2$ and $\Pm$ are the same.  What is
more, we are concerned in this work with weakly magnetized plasmas,
$\beta>1$, and generally $\beta\gg1$.  Therefore, at the threshold
$\Pm=1$, the plasma is never dilute, and the collisional regime is valid.
Note, however, that once into the large $\Pm$ regime, substantial heating and
magnetic field growth may lead to a dilute plasma phase, or perhaps
even to a fully collisionless phase in which the fluid approximation
itself breaks down.

\section{Analysis}

Our goal is a simple one: we wish to follow the behavior of $\Pm$ with
disk radius in a standard $\alpha$ model, in effect testing such models
for self-consistency.  If most of the energy extracted from differential
rotation is locally dissipated, the basic $\alpha$ scalings are probably
robust.  This is particularly true if the problem is framed to minimize
any possible explicit dependence upon $\alpha$ of the temperature and
density, as we have done.  Then, even if in real disks it is not a very
good approximation to treat $\alpha$ as a constant, its variability is
not crucially important for the scaling laws.

\subsection {Pm behavior in $\alpha$ models}

Our starting point is the Kramers opacity disk model
of Frank, King, and Raine (2002).
The density in the midplane is 
\beq
\rho = 3.1\times 10^{-8} \alpha^{-7/10} {\dot M_{16}}^{11/20}(M/M_\odot)^{5/8}
{R_{10}}^{-15/8} q^{11/20}\ {\rm gm}\, {\rm cm}^{-3},
\eeq
where ${\dot M_{16}}$ is the mass accretion rate in units of $10^{16}$
g s$^{-1}$, $M/M_\odot$ is the central mass in solar units, $R_{10}$ is
the cylindrical radius $R$ in units of $10^{10}$ cm., and $q=1-(R_*/R)^{1/2}$.
The quantity $R_*$ is a fiducial radius at which the stress is taken to vanish
(the ``inner edge''), but in practice we shall assume that $R \gg R_*$, and
hence that $q$ is unity.  The midplane temperature is given by Frank et al. (2002)
as
\beq
T= 1.4\times 10^{4} \alpha^{-1/5} {\dot M_{16}}^{3/10}(M/M_\odot)^{1/4}
{R_{10}}^{-3/4} q^{3/10}\ {\rm K}.
\eeq
This leads to a Prandtl number of
\beq
\Pm = 
9.0\times 10^{-5}
l^{-1} \alpha^{1/10} {\dot M_{16}}^{13/20} (M/M_\odot)^{3/8} {R_{10}}^{-9/8} q^{13/20}.
\eeq
Typical disk Prandtl numbers are therefore very small, and insensitive
to scaling with $\alpha$.  Transitions from low to high $\Pm$, if they occur at all,
will occur in the inner disk regions.

Let us calculate $R_{cr}$, the critical radius at which $\Pm=1$.  Here, it will suffice
to set $q=1$ ($R\gg R_*$); a more accurate numerical calculation (described below)
certainly justifies this.   With $\Pm=1$, we find
\beq
R_{cr} = 2.5\times 10^{6}\  l^{-8/9} \alpha^{-4/45} {\dot M_{16}}^{26/45} (M/M_\odot)^{1/3}
\ {\rm cm}.
\eeq
The region of interest is evidently on scales of tens of Schwarzschild radii ($R_S$).
With $R_S=2GM/c^2$, this becomes
\beq
{R_{cr}\over R_S} = 8.5 \alpha^{-4/45} {\dot M_{16}}^{26/45} (M/M_\odot)^{-2/3} l^{-8/9}.
\eeq
Our final step is to scale the mass accretion rate with $M$.  If we
assume that the source luminosity $L$ is a fraction $\epsilon$
of ${\dot M}c^2$ and a fraction $\delta$ of the Eddington luminosity
$$L_{Edd}=1.26\times10^{38} (M/M_\odot)\ {\rm erg}\ {\rm s}^{-1},$$ then
\beq\label{rcrit}
{R_{cr}\over R_S}=59 \left(\alpha_{-2} M/M_\odot\right)^{-4/45} (\delta/\epsilon)^{26/45}
l^{-8/9}
\eeq
where $\alpha_{-2}$ is $\alpha$ in units of 0.01.  The ratio
$\delta/\epsilon$ is just the mass accretion rate measured in units
of the Eddington value $\dot{M}_{Edd}=L_{Edd}/c^2$.   This shows that the critical
radius at which the Prandtl number transition occurs, when measured in
units of $R_S$, is remarkably insensitive to the central mass.  In general
we find that
$R_{cr}$ varies roughly between 10 and 100 $R_S$.  In principle, the low
$\Pm$ region could in some cases extend all the way down to $2-3 R_S$,
particularly for larger AGN masses.  Iron line observations of, for
example, the well-studied Seyfert galaxy MCG--60--30--15, (Fabian et
al. 2002) suggest the presence of an ordinary Keplerian-like disk
down to $3R_S$, and the $\Pm$ transition hypothesis must accommodate this:
no transition should also be a possibility.

\subsection{$\Pm$ behavior in numerical $\alpha$ models}

The result of the previous section neglects radiation pressure and
electron scattering contributions to the opacity.  In
particular, the radiation to gas pressure ratio
is easily calculated.  With $q=1$,
\beq
{P_{rad}\over P_{gas}} = 5\times 10^{-3} 
\alpha^{1/10} {\dot M_{16}}^{7/20}(M/M_\odot)^{1/8}
{R_{10}}^{-3/8}.
\eeq
(This differs from equation [5.56] in Frank et al. [2002].)
At $R=R_{cr}$,
\beq
{P_{rad}\over P_{gas}}=
0.16 l^{1/3} \left( {\alpha\delta\over\epsilon}{M\over M_\odot}
\right)^{2/15}.
\eeq
This varies between a $10\%$ and an order unity effect for applications
of interest.  To ensure that radiative corrections do not alter the basic
conclusion of the existence of a crtical $\Pm$ transition radius under
nominal conditions, we have adapted the disk
code of Terquem \& Papaloizou (1999) to construct more detailed $\alpha$
models.  Both radiation pressure and electron scattering opacity were
included.  We find that the essential qualitative features of equation
(\ref{rcrit}) remain intact, though radiative effects do alter the
scalings somewhat.  We focus on two central masses, one a source of $10
M_\odot$ (representative of an X-ray binary), the other $10^8 M_\odot$,
which is representative of an AGN.  The Prandtl number behavior for
each of these cases for several different values of $\alpha$, but at
a fixed accretion rate ($0.1\dot{M}_{Edd}$), is shown in figure (1).
The two cases are very similar.  Starting with a standard Keplerian
$\alpha$ disk, these black hole accretion sources seem to make a transition
from low $\Pm$ to high $\Pm$ at a typical value of $\sim 50 R_S$.

Figures (2) and (3) show $\Pm$ plots as meridional slices.  A central
mass of $10M_\odot$ is assumed for figure (2), while figure (3)
corresponds to $10^8 M_\odot$.  In each figure, the left and right
diagrams correspond respectively to $\dot M/ \dot{M}_{Edd} =0.001,
1$.  We have used $\alpha=0.05$.  At higher accretion rates, the $\Pm>1$
region can be extensive; on the other hand, if ${\dot M}/\dot{M}_{Edd}$
is sufficiently small, the flow can have $\Pm<1$ down to the marginally
stable orbit $R=3R_S$.

%10SolarMasses_alpha.eps   
%1e8SolarMasses_alpha.eps

\begin{figure}
%\plottwo{Alpha10M.eps}{Alpha1e8M.eps}
\plottwo{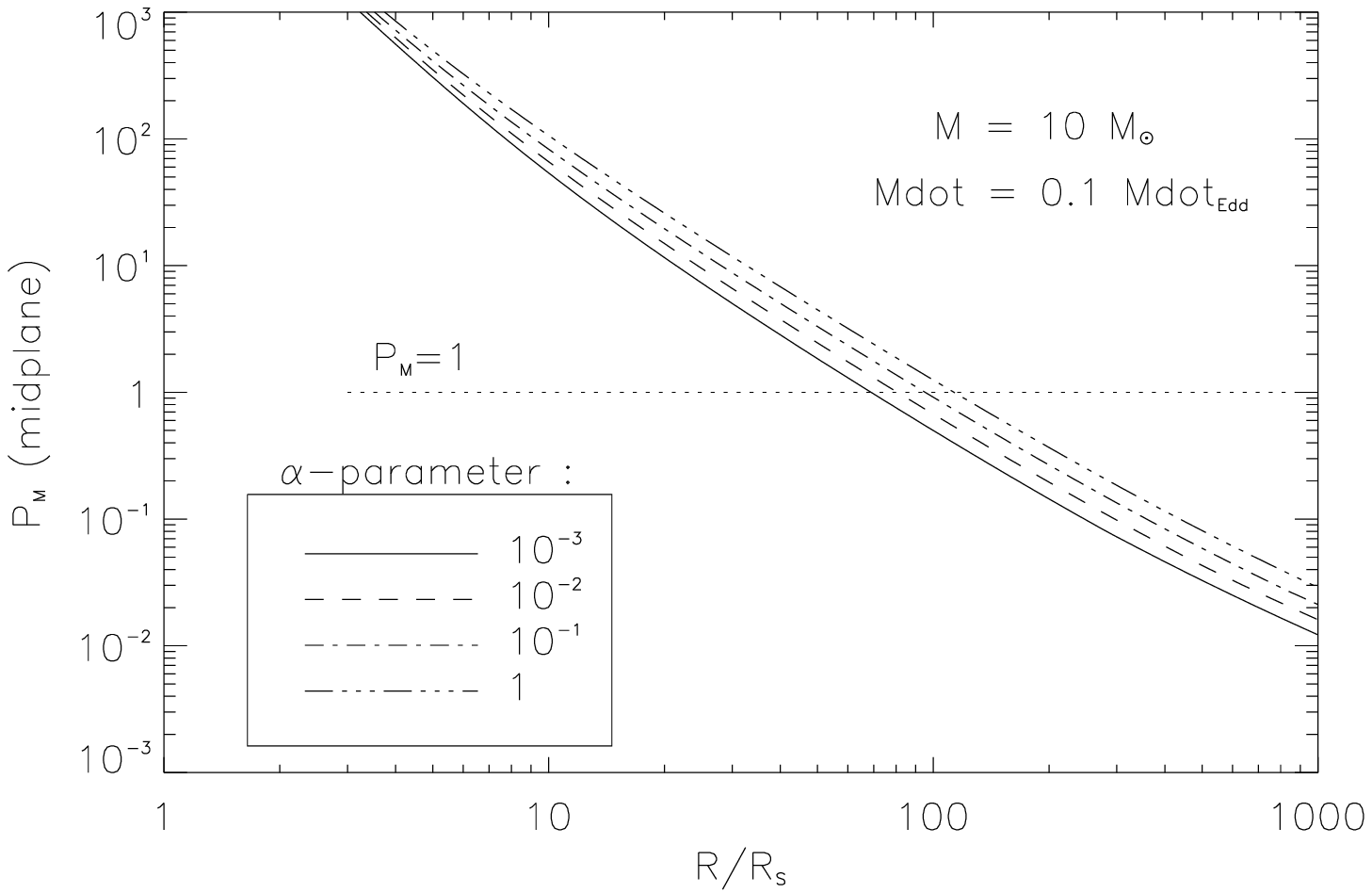}{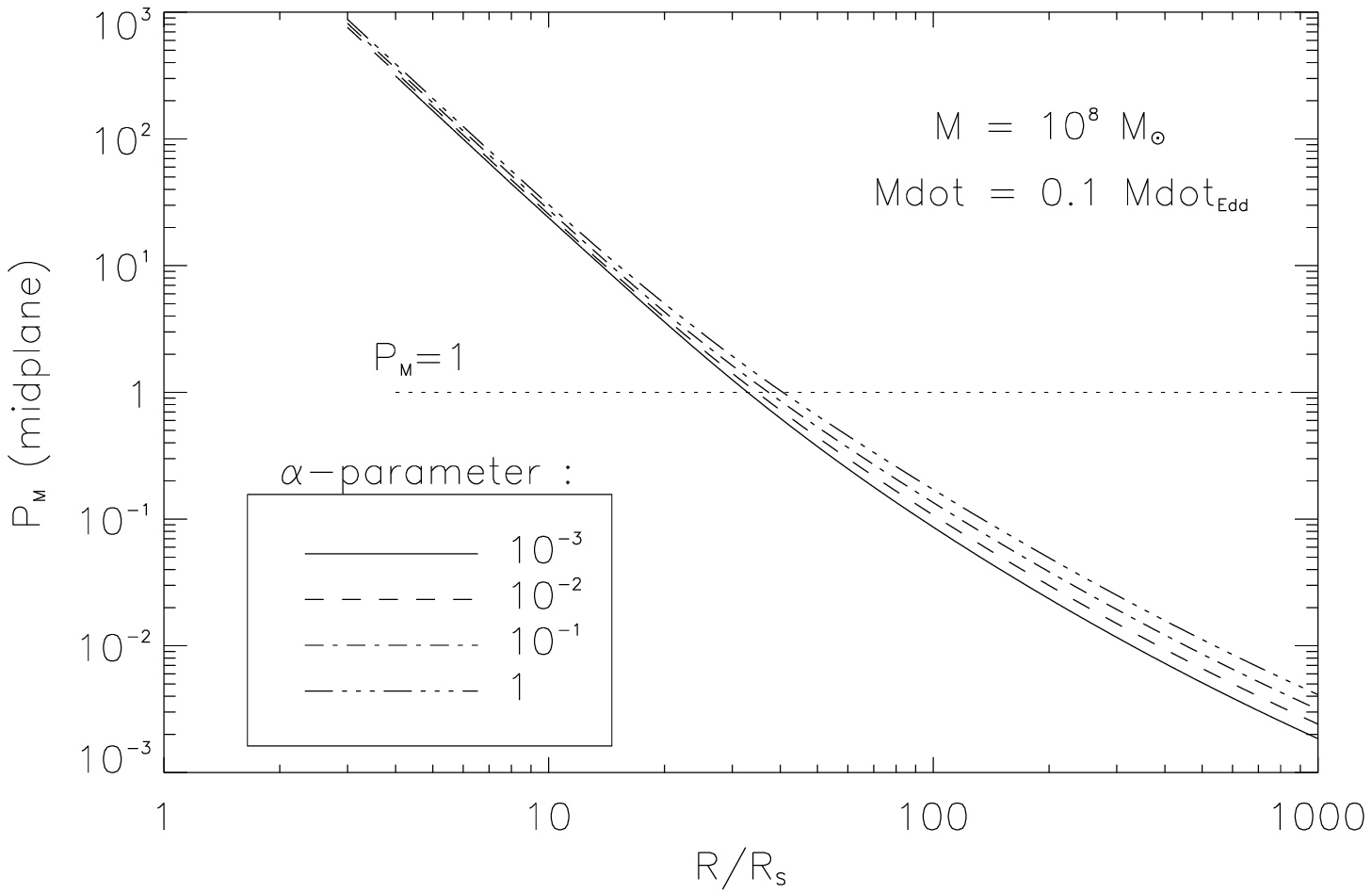} 
\caption{Behavior of $\Pm$ for $10 M_\odot$ (left) and $10^8 M_\odot$
(right) black holes for several different $\alpha$ values as a function
of disk radius $R/R_S$, where $R_S$ is the Schwarzschild radius.
The accretion rate is taken to be $0.1\dot{M}_{Edd}$.  Calculations were
carried out using the code of Terquem \& Papaloizou (1999) including
radiation pressure and electron scattering corrections to the opacity.
Note the insensitivity of the results to both $\alpha$ and the central
mass.} \end{figure}

\begin{figure}
\plottwo{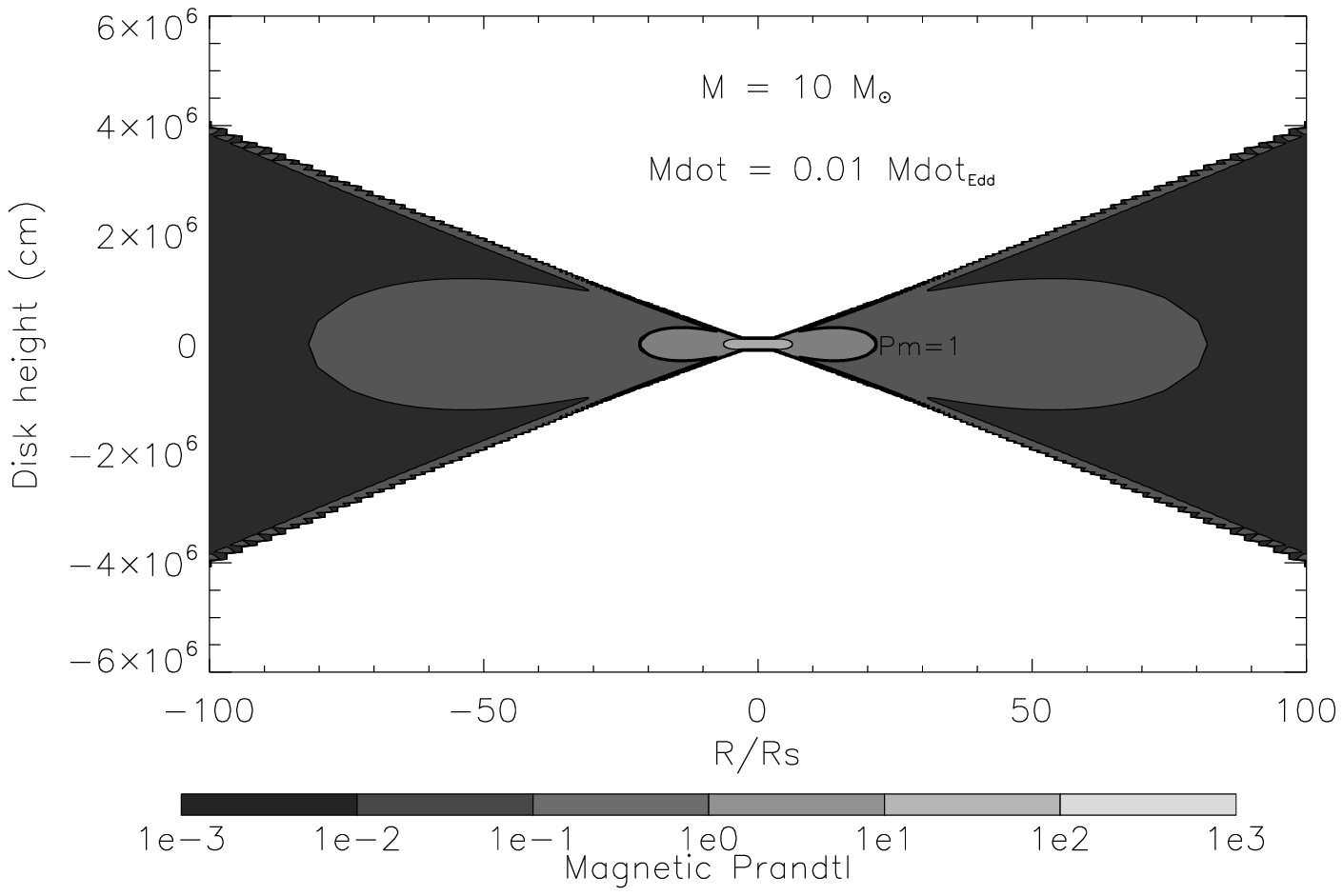}{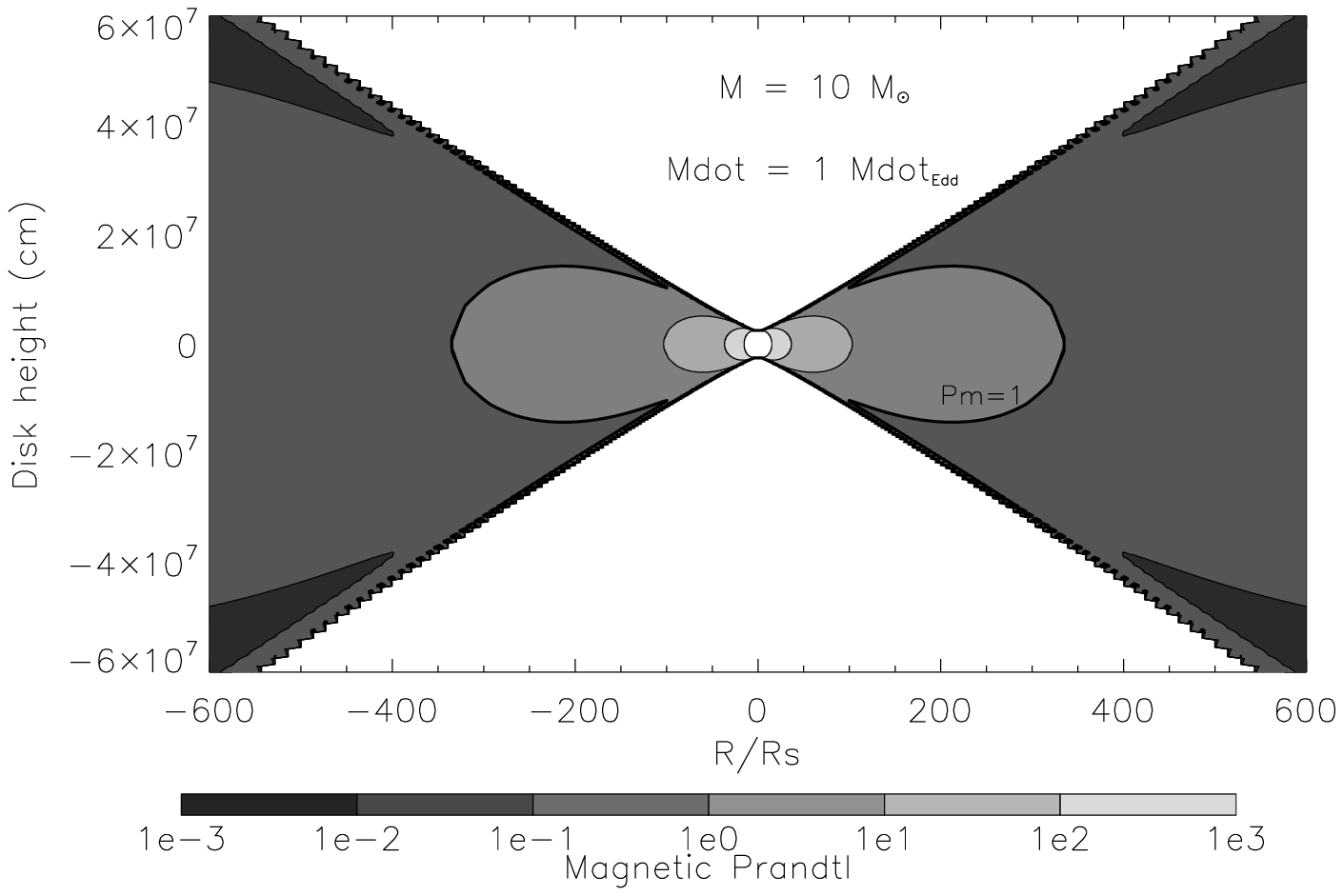}
\caption{Regions of $\Pm$ for a $10M_\odot$ black hole accretion disk.
Left diagram corresponds to $\dot M=0.01\dot{M}_{Edd}$, right
to $\dot M= \dot{M}_{Edd}$. In both cases, $\alpha=0.05$.  At large accretion rates, the high
$\Pm$ regions can be quite extended.  (Note change of radial scale.)}
\end{figure}

\begin{figure}
\plottwo{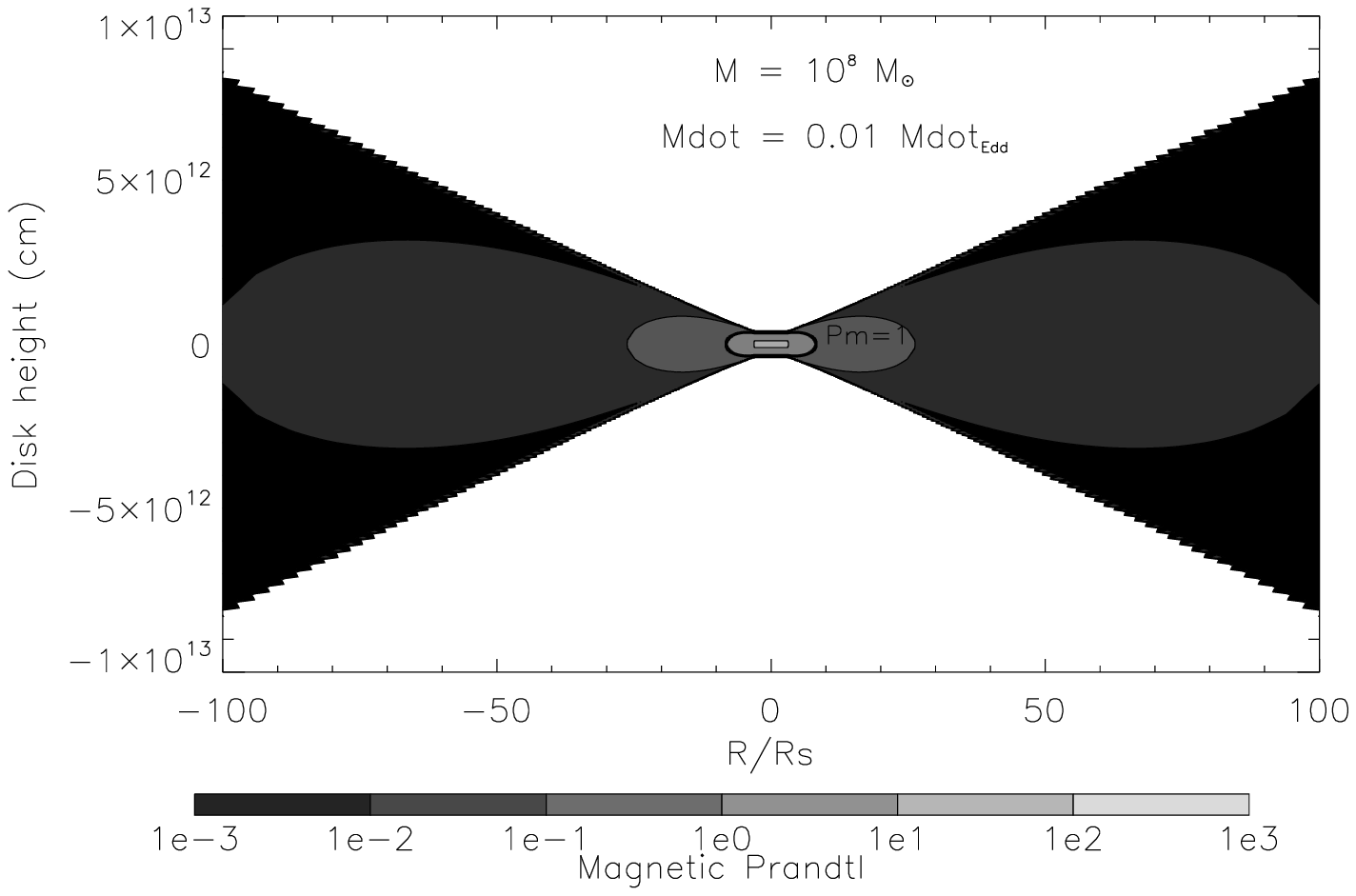}{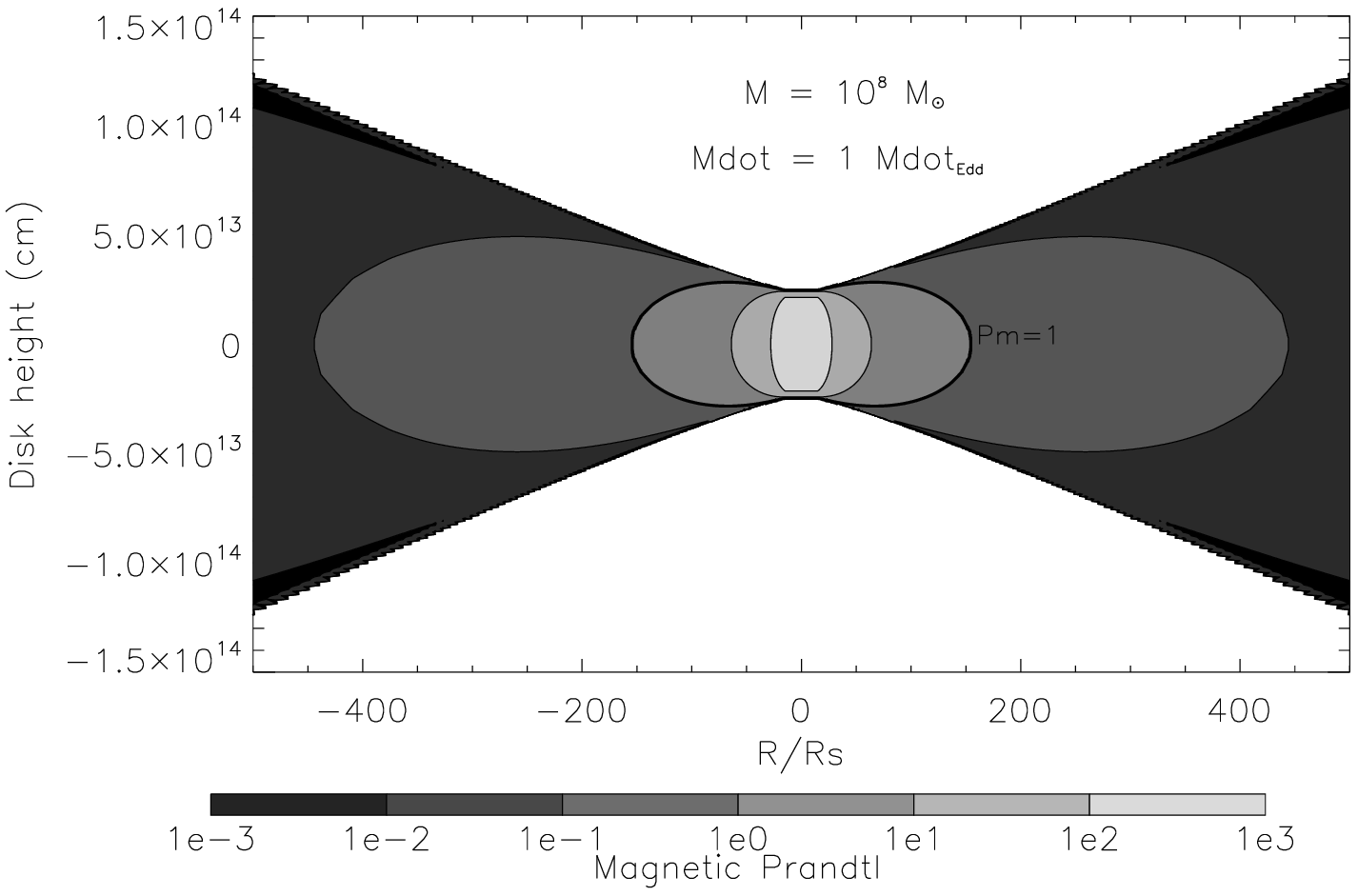}
\caption{Regions of $\Pm$ for a $10^8 M_\odot$ black hole accretion disk.
Left diagram corresponds to $\dot M=0.01\dot{M}_{Edd}$, right
to $\dot M= \dot{M}_{Edd}$. In both cases, $\alpha=0.05$. 
The high $\Pm$ regime is very limited in the low $\dot M$ case 
on the left. (Note change of radial scale.)}
\end{figure}

%10M_deltaepsilon001.eps  10SolarMasses_alpha.eps   1e8M_deltaepsilon1.eps
%10M_deltaepsilon01.eps   1e8M_deltaepsilon001.eps  1e8SolarMasses_alpha.eps
%10M_deltaepsilon1.eps    1e8M_deltaepsilon01.eps

\section {Discussion}

The findings of the previous section show that if $\alpha$ models
are even qualitatively correct in their scalings, only black holes
and neutron star accretion disks, i.e., classical X-ray sources, will
have regions with $\Pm<1$ and $\Pm>1$.  If, as we would argue, there
is a physical difference in the saturated state of MHD turbulence in
these two regimes, it should be reflected in the astrophysical behavior
manifested by X-ray sources.  We tentatively suggest that the principal
hard high states and low soft states associated with these sources is
related respectively to the relative radiative dominance of the $\Pm>1$
and $\Pm<1$ accretion regions.  In this discussion we will outline
arguments that are suggestive, but as yet far from conclusive, of this.
They are meant to spur further numerical investigation in what could
prove to be an interesting direction.

The results of several independent numerical simulations of MHD
turbulence, both forced and shear-driven, appear to indicate that
if $\Pm>1$, field dissipation becomes more inefficient, apparently
because viscous stresses make the resistive scale less accessible
(Fromang et al. 2007, Lesur \& Longaretti 2007, Iskakov et al. 2007).
If field dissipation is inefficient, the most likely scenario is that
the field will initially build up on the viscous scale, but ultimately
cascade upward to larger scales \citep{b01}.  In a disk, the growing
magnetic field would drive the MRI more vigorously until ultimately ---
and ``ultimately'' may in fact be rather rapid  --- the field is of order
thermal.  At this stage further MRI development is likely to be halted.

The effective absence of resistivity of course does not mean that
dissipation is absent; dissipative heating will still be present in the form
of viscous heating.  Note that the dominance of the resistive scale in $\Pm<1$
turbulence means that the electrons are directly heated (assuming that
classical Spitzer resistivity applies), whereas the ions are directly
heated in viscosity dominated $\Pm>1$ turbulence.  The need for the
dominance of ion heating in low luminosity black hole accretion is
by no means a new idea (e.g. Narayan \& Yi 1995), but placing it within
the Prandtl number framework lends mutual support to this current work
and to what has become the standard picture.   In addition, the heating of
a $\Pm>1$ magnetized plasma may be very vigorous---unlike ohmic
resistivity, viscous thermalization 
does not destroy the current sources. 

Conditions for a thermal runaway are present: at constant pressure, $\Pm
\propto T^5$.  Since $\Pm$ is an {\em increasing} function of temperature,
a little heating would tip $\Pm\sim 1$ accretion towards the direction
of $\Pm \gg 1$ accretion.  This would mean yet greater heating, following
the numerical lead
that large Pm turbulence is characterized by greater fluctuation levels.
But this argument works in both directions, cooling as well as heating.
A formal boundary between high and low Prandlt number regions would,
in this view, be unstable because of the dependence of Pm upon $T$.
This can be investigated by direct simulation.  We speculate that the
$\Pm<1$ region is a standard $\alpha$ disk and responsible for soft
thermal emission; the $\Pm>1$ region corresponds to lower density,
hotter accretion.  Although such a component has been regarded as
essential for understanding the X-ray spectra of black hole accretion
sources, the reason that a disk would suddenly make the transition from
one type of flow to the other has always been unclear.  Prandtl number
considerations may make this transition less mysterious.

The unstable boundary between high and low Prandtl number flow marks a
fundamental change in the accretion flow, leading to a distinct inner
accretion zone that dominates the hard tail of the X-ray spectrum.
A better understanding of the interface will help to establish whether it
is involved with transitions from one state to another.  In a subsequent
publication, we will present a technique to make this problem tractable
and predictive.

At this point the major gap in our scenario is the difference between
the modest but rigorous numerical findings of a correlation between
$\Pm$ with the amplitude of the turbulent stress, and the full blown
thermal runaway that we envisage.  That gap can begin to be filled
with well-crafted numerical investigations of temperature-dependent
dissipation coefficients in MRI turbulence.   Little has yet been
done along these lines, and it promises to be extremely challenging,
particularly if the ideas presented here are qualitatively correct and a
dilute or collisionless plasma appears.  But preliminary investigations
have already begun.

We end by noting that in the paper introducing the MRI to the
astrophysical community (Balbus \& Hawley 1991), two possible nonlinear
resolutions were envisioned.  In one the field was limited to subthermal
strengths by vigorous dissipation; in the other it grew to thermal
levels and became ``stiff.''  Subsequent numerical simulations seem to
support the first outcome, but this may well have been because the codes
used were not in the large $\Pm$ regime.  Both scenarios might in fact
be equally viable, the choice of direction being made by the Prandtl
number of the turbulence.

\section*{Acknowledgements.}  It is a pleasure to thank Alex Schekochihin
for valuable conversations on high $\Pm$ turbulence, as well as Julian
Krolik, Jim Stone, and an anonymous referee for detailed comments
that improved this manuscript.  This work was supported by a Chaire
d'Excellence award to S. Balbus from the French Ministry of Higher
Education, by NASA grants NNG04GK77G and NAG5-13288, and by
NSF grant PHY-0205155

\section* {Appendix: Collision time and viscosity estimates.}

Spitzer (1962) defines a ``deflection time'' $t_D$ for a test particle
(denoted by subscript $t$) of mass $m_t$, charge $Z_t$ (in units of $e$),
and velocity $w_t$
to be scattered by Coulomb interactions
by a population of field particles
(denoted by subscript $f$) of number density $n_f$.
It is given by 
\beq
t_D(t\rightarrow f) = {m_t^2 w_t^3 \over
8\pi n_f e^4 Z_t^2 Z_f^2\,  F(l_fw_t)\, \ln\Lambda_{tf}}.
\eeq
Here,
\beq
l_f =\sqrt{m_f\over 2kT_f}
\eeq
with $m_f$ and $T_f$ refering respectively to the mass and temperature
of the field particles.  The function $F(x)$ is
\beq
F(x) = 
\left(1 -{1\over 2x^2}\right) {\rm erf}(x)+
{e^{-x^2}\over x \sqrt{\pi }}
\eeq
where erf$(x)$ denotes the standard error function
\beq
{\rm erf}(x) ={2\over\sqrt{\pi}}\int_0^x e^{-s^2}\, ds.
\eeq
The argument of the logarithm is
\beq
\Lambda_{tf}= {1.5\over Z_tZ_f e^3} \, \left( k^3 T_t^2 T_e\over \pi n_e
\right)^{1/2}
\eeq
where $T_t$ and $T_e$ refer to the test particle and electron temperature,
respecitvely, and $n_e$ is the electron density.  When the test particles
are electrons, then for temperatures in excess of $4.2\times 10^5$K, an
additional factor of $(4.2\times 10^5/T_e)^{1/2}$ appears in the expression
for $\Lambda$ (a correction for quantum diffraction).

In what follows, we shall always take a single temperature ($T$) fluid,
and set $w_t$ equal to the rms test particle velocity, i.e.,
$m_tw_t^2=3kT$.  Then,
\beq
t_D(t\rightarrow f) = { m_t^{1/2} (3kT)^{3/2} \over
8\pi n_f e^4 Z_t^2 Z_f^2\,  F(\sqrt{3m_f/2m_t})\, \ln\Lambda_{tf}}.
\eeq
and, with $n_H$ denoting hydrogen number density,
\beq
\Lambda_{tf}= {1.5\over Z_tZ_f e^3} \, \left( k^3 T^3 \over \pi n_e
\right)^{1/2}= {1.239\times 10^4\over Z_t Z_f} {T^{3/2}\over n_e^{1/2} }
 = {1.131\times 10^4\over Z_t Z_f} {T^{3/2}\over n_H^{1/2} }
 \eeq
with the additional diffraction correction of a factor
of $(4.2\times 10^5/T)^{1/2}$ needed for the case of electron test
particles as noted above.  For our cosmic gas, $n_e=1.2n_H$ under
the assumption of fully ionized helium.
If a single temperature prevails, then
$\Lambda_{tf}=\Lambda_{ft}$; note that the time
$t_D(t\rightarrow f)$ does not have a similar
symmetry between $t$ and $f$.  

As discussed in the text, representative values for $T$ and $n_H$
near the Prandtl number transition are $T=6\times 10^6$ K and
$n_H=2\times 10^{22}$ cm$^{-3}$.  For these values,
\beq
\ln\Lambda_{\rm H\,H}=7.07, \qquad 
\ln\Lambda_{\rm H\, He}=6.38,               
\eeq
and
\beq
t_D(H\rightarrow H) = 1.614\left(T^{3/2}\over n_H\right)\left(7.07\over\ln\Lambda_{\rm H\,H}
\right)\ {\rm s.}
\eeq

The dynamical ion viscosity $\eta_V$ of a fully ionized plasma is 
(Spitzer 1962):
\beq
\eta_V= 
{0.406 m^{1/2} (kT)^{5/2} \over Z^4 e^4\, \ln \Lambda}\ 
{\rm g}\, {\rm cm}^{-1} \ s^{-1}
\eeq
where both the test and field particles are identified with ions
of mass $m$ and charge $Z$.  Dimensionally, this takes the form
\beq
\eta_V = C_{\eta_V} \ \rho w^2 t_D
\eeq
where $C_{\eta_V}$ is a numerical constant (nominally but universally
1/3 in elementary modeling),
$\rho$ is the ion density, $w^2$ is the mean squared ion thermal
velocity ($3kT$ divided by the ion mass), and $t_D$ is the ion-ion deflection time.  
In considering a cosmic mixture of a 10\% helium abundance
fraction, one must take into account modifications to $t_D$ due
to scattering of protons by He nuclei, in addition to the contribution
to the viscous stress carried by these nuclei.  Because of
the sensitive dependence on atomic number $Z$, a relatively small amount
of He could in principle make a significant contribution to $\eta_V$.
Indeed, fully ionized metals at the level of a few per cent
also make a contribution 
because of the $Z$ scaling, but we shall ignore this here.
Assuming that $C_{\eta_V}$
is the same for all species, an estimate for the cosmic abundance
viscosity is then  
\beq
\eta ({\rm cosmic})= C_\eta\left(\rho_H w_H^2 t_D(H) + \rho_{He}
w_{He}^2 t_D(He)\right)
\eeq
where the deflection times are now given by
\beq
{1\over t_D(H)} = 
{1\over t_D(H\rightarrow H)}
+{1\over t_D(H\rightarrow He)}=
{1\over t_D(H\rightarrow H)}\left( 1 + {t_D(H\rightarrow H)\over 
t_D(H\rightarrow He)} \right),
\eeq
\beq
{1\over t_D(He)} = {1\over t_D(He\rightarrow H)}
+{1\over t_D(He\rightarrow He)} =
{1\over t_D(He\rightarrow H)}\left( 1 + {t_D(He\rightarrow H)\over
t_D(He\rightarrow He)} \right).
\eeq
Now,
\beq
{t_D(H\rightarrow H)\over t_D(H\rightarrow He)} = 
\left(n_{He}\over n_H\right)\times 4\times
\left( F(\sqrt{6})\over F(\sqrt{1.5})\right)
\times \left( \ln\Lambda_{\rm H\, He}
\over \ln \Lambda_{\rm H\, H}\right)
\eeq
and
\beq
{t_D(He\rightarrow H)\over t_D(He\rightarrow He)} =
\left(n_{He}\over n_H\right)\times 4\times
\left( F(\sqrt{1.5})\over F(\sqrt{.375})\right)
\times \left( \ln\Lambda_{\rm He\, He}
\over \ln \Lambda_{\rm He\, H}\right)
\eeq
In each of the above, the ratio of the Coulomb logarithms is about 0.9
across a wide range of densities and temperatures.  Adopting this value,
we find
\beq
{1\over t_D(H)} ={1.46\over t_D(H\rightarrow H)}, \quad
{1\over t_D(He)} ={1.6\over t_D(He\rightarrow H)}.
\eeq
In other words, the effects of test particles interacting with the 10\%
admixture of He results in roughly a 50\% increase in the effective collision
rate.  
At the level of accuracy with which we are concerned, we shall
a deflection time shortening factor of 2/3 in both cases.
The effective viscosity is then
\beq
{2\over 3}C_\eta\left[\rho_H w_{H}^2 t_D(H\rightarrow H) +\rho_{He} w_{He}^2
t_D(He\rightarrow H)
\right]= {2\over3}\rho\nu_H\left[ 1 + {\rho_{He}\over \rho_H}\times{w_{He}^2\over
w_{H}^2}\times {t_D(He\rightarrow H)\over t_D(H\rightarrow H )}\right]
\eeq
where $\rho \nu_H$ is the dynamical viscosity in a gas of pure hydrogen
($\nu$ being the corresponding kinematic viscosity).
The final deflection time ratio is
\beq
{t_D(He\rightarrow H)\over t_D(H\rightarrow H )}=2\times(1/4)\times [F(\sqrt{1.5})/
F(\sqrt{.375})]\times [\ln\Lambda_{\rm H\,H }/\ln\Lambda_{\rm He\, H}]\simeq0.925
\eeq
The final estimate for the cosmic abundance viscosity is
\beq
\rho\nu({\rm cosmic}) = {2\over3}\times 1.09\times\rho  \nu_H = 1.6\times 10^{-15} {T^{5/2}
\over \ln\Lambda_{\rm H\, H}} {\rm gm}\, {\rm cm}^{-1}\, {\rm s}^{-1}
\eeq

\end{document}